\documentclass[preprintnumbers,aps,twocolumn,floatfix,superscriptaddress]{revtex4-1}
\usepackage{eso-pic,calc}
\usepackage{graphicx}
\usepackage{amsmath, amsthm, amssymb}
\usepackage{epsfig}
\usepackage{bm}
\usepackage{color}

\usepackage[colorlinks=True]{hyperref}  
\hypersetup{
 	colorlinks=true,  
 	linkcolor=blue,  
 	citecolor=blue, 
 	filecolor=magenta,   
 	urlcolor=blue         
 }
 
\def\beqr{\begin{eqnarray}}
\def\eqnr{\end{eqnarray}}
\def\beq{\begin{equation}}
\def\bc{\begin{center}}
\def\ec{\end{center}}
\def\eqn{\end{equation}}

\def\etl{$et~al.$}

\begin{document}
\title{Scaling behavior in the number theoretic division model of self-organized criticality}
\author{Rahul Chhimpa}
\affiliation{Department of Physics, Institute of Science,  Banaras Hindu University, Varanasi 221 005, India}

\author{Avinash Chand Yadav\footnote{jnu.avinash@gmail.com}}
\affiliation{Department of Physics, Institute of Science,  Banaras Hindu University, Varanasi 221 005, India}

\begin{abstract}
{We revisit the number theoretic division model of self-organized criticality [\href{https://journals.aps.org/prl/abstract/10.1103/PhysRevLett.101.158702}{Phys. Rev. Lett. 101, 158702 (2008)}]. The model consists of a pool of $M-1$ ordered integers $\{2, 3, \cdots, M\}$, and the aim is to dynamically form a primitive set of integers, where no number can be divided or divisible by others. Using intensive simulation studies and finite-size scaling method, we find the primitive set size fluctuations in the division model to show power spectral density of the form $1/f^{\alpha}$ in the frequency regime $1/M\ll f \ll 1/2$ with $\alpha \approx 2$ (different from $\alpha \approx 1.80(1)$ as reported previously) along with an additional scaling in terms of the system size $\sim M^b$. We also show similar power spectra properties for a class of random walks with a power-law distributed jump size (L\'evy flights).  }
\end{abstract}
\maketitle

\section{Introduction}
Equilibrium systems undergoing continuous phase transition demonstrate scaling behavior under a very restrictive condition (near a critical point). However, many non-equilibrium systems naturally evolve towards a self-organized critical (SOC) state~\cite{PhysRevLett.59.381, Jensen_1998, PhysRevE.60.2706, Pruessner_2012, MARKOVIC201441, RevModPhys.90.031001}, marked by the presence of long-range space-time correlations, usually characterized as \emph{fractal} structures or/and \emph{$1/f$ noise}~\cite{RevModPhys.53.497, pradhan_2021, PhysRevE.110.014106}. Past studies revealed numerous systems exhibiting the SOC phenomenon, ranging from sandpile~\cite{PhysRevLett.59.381, DHAR200629} and earthquakes~\cite{PhysRevLett.88.178501, Nekrasova_2020, Perinelli_2024} to species evolution~\cite{PhysRevLett.71.4083, PhysRevE.108.044109} and neuronal activity~\cite{PhysRevLett.96.028107, levina_2007, PhysRevX.9.021062}. The interplay between slow drive and instantaneous dissipation (separation of time scales) is a typical characteristic that results in the system being in a critical state, where the response to an external drive is nonlinear.

Number theory deals with numbers and their relationships and offers interesting links with physics. An example includes the Reimann zeta function related to a partition function~\cite{julia_1990}. To investigate the emergence of SOC from the perspective of number theory, Luque \etl~\cite{PhysRevLett.101.158702} introduced a pertinent \emph{number-theoretic division model} (details discussed below). The division model consists of a pool, a set of ordered integers $\{2, 3, \cdots, M\}$ with $M-1$ elements. The objective of the model is to dynamically achieve a primitive set of integers, where no number can be divided or divisible by others. The elementary steps of the dynamics are the following: A randomly selected integer from the pool is introduced in the primitive set, and if the new number divides and is divisible by $s$ existing numbers, then these $s$ integers are put back in the pool and termed as a division avalanche of size $s$. The same dynamical update occurs iteratively.

The simple division model exhibits intriguing scaling features. One of the quantities of interest in the model is the fluctuations in the primitive set size $N(t)$. The primitive set is empty initially, and its size changes randomly with the model dynamics. It effectively increases during a transient regime; eventually, it hovers around a mean value that varies with the system size as $\sim M/\ln{M}$. To understand the temporal correlations of the primitive set size, Luque \etl~\cite{PhysRevLett.101.158702} studied the power spectral density and found it to decay in a power law in terms of frequency with an exponent of $1.80(1)$ (a clear signature of $1/f$ noise or the presence of non-trivial temporal correlation). 
Since the division avalanche of size $s$ changes the primitive set size, the avalanche size probability distribution is also a relevant quantity. Interestingly, the division avalanche size distribution follows a power-law decay below a cutoff with an exponent of 2.0(1)~\cite{PhysRevLett.101.158702}. The cutoff size grows nonlinearly with the system size as $\sim M/\ln{M}$.

However, the temporal correlation in the primitive set size remains poorly understood, as discussed below. 
In this paper, we numerically present a finite-size scaling (FSS) analysis of the power spectra of $N(t)$ in the division model and reveal $1/f^2$-type scaling in the frequency regime $1/M\ll f \ll 1/2$ (slightly different from $1/f^{1.8}$ as reported previously), along with an intriguing system size scaling $\sim M^b$ with $b \approx 0.77$. We further show power spectra for a class of one-dimensional (1D) random walks with a power-law distributed jump and observe that the temporal correlation in the random walk behaves quite similarly to that of the primitive set size noise.

We emphasize that the power spectra of a noisy process in SOC systems can show system size scaling in the non-trivial frequency regime. The temporal variation in local quantity (in a sandpile on a narrow strip~\cite{Yadav_2022}, a continuous state sandpile~\cite{Kumar_2022}, and a train model~\cite{PhysRevE.104.064132}) typically shows $1/f^2$-type scaling along with a system size scaling $L^b$ with $b<0$. However, the local fitness in the original Bak-Sneppen model follows $1/f^{1.2}$ and $ 1/L$ type of scaling features~\cite{PhysRevE.108.044109}. On the other hand, the system size scaling in the same frequency regime is typically absent or may be present for a global noisy process. Examples include the configuration dynamics in a sandpile or the position of a random walk on a ring, which can show scaling features of the form $1/f^{3/2}$ and $\sim L$~\cite{Yadav_2022}. 

The subsequent Sec.~\ref{sec_2} begins with recalling the division model and its salient features. In Sec.~\ref{sec_3}, we present our numerical results for the primitive set size power spectra analyzed by the finite-size scaling, uncovering intriguing scaling behaviors. In Sec.~\ref{sec_4}, we present a random walk with a power-law distributed jump and its power spectra properties, showing that the process behaves similarly to that of the primitive set noise. Finally, Sec.~\ref{sec_5} presents a summary of the paper. 

\section{Number theoretic division model}\label{sec_2}
As proposed by Luque \etl~\cite{PhysRevLett.101.158702}, the definition of the division model that exhibits SOC is the following: Consider an ordered set $\{2, 3, \cdots, M\}$, with $M-1$ integers (a pool). The pool does not include 0 and 1, and no integers repeat. With the aid of the pool, the goal is to form dynamically another set of integers, namely, the primitive set, which has $N(t)$ numbers of integers at time $t$. The model dynamics consist of two main evolution steps: external drive and dissipation. Randomly selecting an integer from the pool and adding it to the primitive set represents the external drive or a small perturbation. If the added integer is one that exactly divides and is divisible by $s_1$ and $s_2$ integers within the primitive set, respectively, a division avalanche of size, $s = s_1 + s_2$, occurs by returning them into the pool (dissipation). In this way, the primitive set eventually restores its primitiveness by changing the size to  
\begin{equation}
N(t+1)=N(t)+1-s(t),\nonumber
\end{equation} 
with an initial condition $N(t=0)=0$.
The model dynamics include a separation of time scales. The dissipation event (the return of integers to the pool) is instantaneous, while the external drive (the addition of integers to the primitive set) is paused until the set becomes primitive again (slow drive).

Luque \etl~\cite{PhysRevLett.101.158702} argued that the primitive set size follows a mean-field equation 
\begin{equation}
N(t+1)=N(t)+1-\left( \frac{2\ln{M}}{M} \right) N(t),\nonumber
\end{equation} 
with a fixed point $N_c\sim M/\ln{M}$, around which the system self-organizes. The insight for the characteristic size $N_c$ can be obtained from the number theory. A divisor function that provides the number of divisors of $n$ excluding 1 and $n$ is
\begin{equation}
d(n) = \sum_{k=2}^{n-1}\left( \Bigl \lfloor{\frac{n}{k}\Bigr \rfloor-\Bigl\lfloor{\frac{n-1}{k}}}\Bigr\rfloor\right),\nonumber
\end{equation}
where $\lfloor \cdot \rfloor$ represents the integer part.
Given an integer between $[2, M]$, the mean number of divisors are $\sum_{n=3}^{M}d(n)/(M-1) \sim \ln{M}$. 
Then, randomly chosen two numbers from the pool are divisible with a mean probability $\sim 2\ln{M}$/M.

The division model dynamics also has an interpretation in terms of complex networks.
Networks can describe complex systems, where a node represents an entity and the link between entities represents an interaction~\cite{RevModPhys.90.031001}. The degree of a node provides information about the number of connections that the node has, reflecting the network's topology~\cite{RevModPhys.74.47}. Components of complex systems often interact with each other in many ways. Since $n$ nodes can have up to $n(n-1)/2$ links in a complete network, we can simplify the system by removing links between weakly interacting or weakly correlated nodes, thereby focusing on the more significant interactions~\cite{PACZUSKI2004158}.
Based on topology and growth, various types of networks are possible, including random~\cite{barabasi_2016}, hierarchical~\cite{10.1073/pnas.1300832110}, small-world~\cite{watts_1998}, and scale-free networks~\cite{doi:10.1126/science.286.5439.509}. Scale-free networks are characterized by a power-law degree distribution, $P(k)\sim k^{-\gamma}$, where $1 < \gamma < 3$. Thus, the network has strong heterogeneity, moving from one node to another. Growth and preferential attachment are the fundamental mechanisms behind scale-free networks~\cite{doi:10.1126/science.286.5439.509}. Several SOC systems, such as the sandpile model~\cite{PhysRevLett.91.148701, PhysRevE.73.046117} and the Bak-Sneppen model~\cite{moreno_2002, diego_2007, PhysRevE.71.057102}, have been studied on scale-free networks to understand the effect of the interplay between topology and dynamics in attaining criticality.

Strikingly, the underlying dynamics of the division model also have a map with complex networks with a scale-free degree distribution~\cite{PhysRevLett.101.158702}. The pool and primitive sets can be considered the network and a subset of the network, respectively. Consider a network with $M$ nodes, where each node can have two possible states (active and silent). The state of nodes evolves via the following algorithm. (i) Perturbation: Randomly select a silent node and make it active. (ii) Dissipation: Connected $s$ nodes in the active state go to the silent state. It represents an avalanche event of size $s$, and the number of active nodes is $N(t)$. The avalanche size distribution follows a power law for a scale-free complex network, implying an underlying scale-free topology induces SOC. 

\begin{figure}[t]
	\centering
	\scalebox{1}{\includegraphics{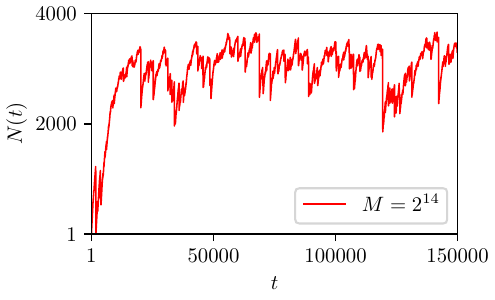}}
	\caption{For the division model, the plot shows a typical realization of the primitive set size, with a pool size of $M = 2^{14}$. The primitive set is initially empty. After the transient period, the signal hovers in time around an average value.}
	\label{fig_dv_ts_1}
\end{figure}

\begin{figure}[t]
	\centering
	\scalebox{1}{\includegraphics{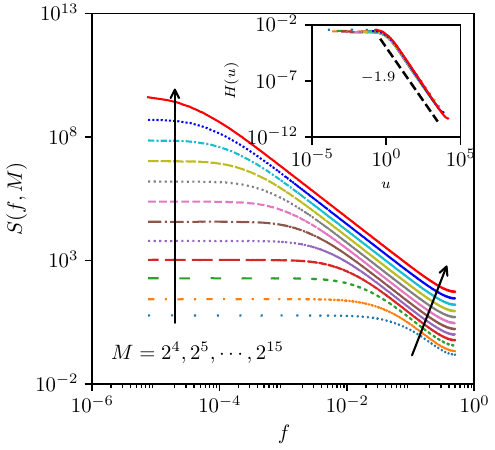}}
	\caption{The division model: (Main panel) The power spectra $S(f, M)$ for the primitive set size fluctuation $N(t)$ with different system size $M$. The arrow marks indicate the increasing trend of the pool size $M$ varying from $2^4$ to $2^{15}$. The signal length is $2^{18}$, and the transient discarded is $10^5$. Each curve is averaged over $10^4$ independent realizations of the signal. (Inset) The scaling function $H(u) \sim S(f, M)/M^{b+\alpha\lambda}$ versus the scaled frequency $u \sim fM^{\lambda}$ [cf. Eqs.~(\ref{eq_ps_1}) and (\ref{eq_ps_2})]. The dashed straight line guides the slope.}
	\label{fig_dv_sf_1}
\end{figure}

\begin{figure}[t]
	\centering
	\scalebox{1}{\includegraphics{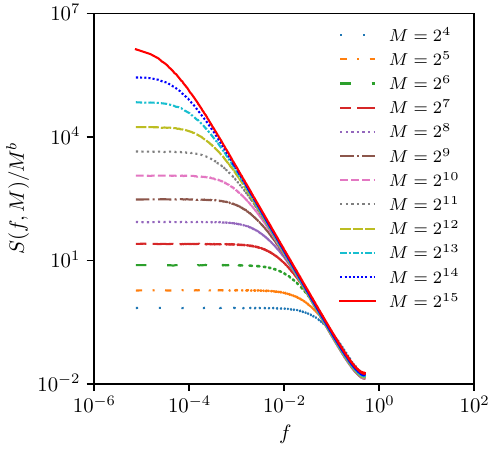}}
	\caption{ The system size dependence of the power spectra in the division model vanishes above the cutoff frequency after plotting the scaled power spectra $S(f, M)/M^b$ with $b = 0.77$.}
	\label{fig_dv_sf_20}
\end{figure}

\begin{figure}[t]
	\centering
	\scalebox{1}{\includegraphics{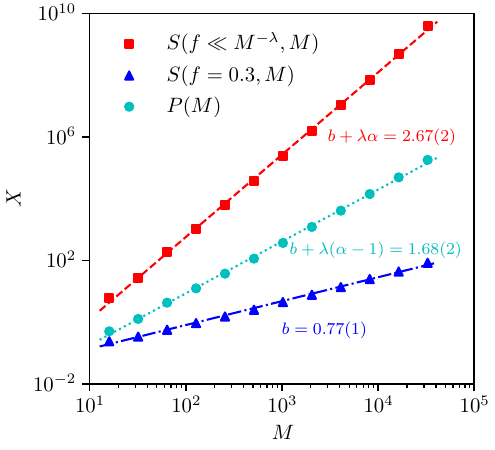}}
	\caption{The division model:  The system size scaling of the power in the low-frequency component $S(f\ll M^{-\lambda}, M) \sim M^{b+\lambda\alpha}$, the power at a fixed frequency above the cutoff frequency $S(f=0.3, M) \sim M^b$, and the total power $P(M) \sim M^{b+\lambda(\alpha-1)}$. The lines are the best-fit plots. }
	\label{fig_dv_sf_2}
\end{figure}

\section{The primitive set size power spectra}\label{sec_3}
Figure~\ref{fig_dv_ts_1} shows a typical realization of the fluctuations in the primitive set size, $N(t)$. To uncover the temporal correlations subtly, we examine the power spectra of $N(t)$ after discarding transients. We used the standard fast Fourier transformation (FFT) method to compute $\tilde{N}(f)$, the Fourier transformation of the signal. The power spectrum is then computed as $S(f) =  \lim_{T\to \infty} \langle |\tilde{N} (f)|^2\rangle/T$, where the angular brackets $\langle \cdot \rangle$ denote ensemble average. Figure~\ref{fig_dv_sf_1} shows scaling behaviors of the power spectra for the fluctuations in $N(t)$ with different system sizes, $M$. The power-law $1/f^{\alpha}$ behavior of the spectrum changes to a flat form below a cutoff frequency $f\ll M^{-\lambda}$. Strikingly, the power spectral density also scales with the system size $\sim M^b$ in the non-trivial frequency regime $M^{-\lambda}\ll f \ll 1/2$ [cf. Fig.~\ref{fig_dv_sf_20}]. As the power spectrum scales with frequency and system size, we can write a scaling ansatz~\cite{PhysRevE.108.044109}
in terms of the reduced frequency $u \sim fM^{\lambda}$ as
\begin{equation}
	S(f,M)  \sim M^{b+\lambda\alpha}H(u),
	\label{eq_ps_1}
\end{equation}
demanding the scaling function $H(u)$ to behave as
\begin{equation}
	H(u) \sim \begin{cases} 1, ~~~~~~~~~~~~{\rm for}~~~~~u\ll 1,\\ 1/u^{\alpha}, ~~~~~~~{\rm for}~~~~~ u \gg1.\end{cases}
		\label{eq_ps_2}
\end{equation}
The power value below the cutoff frequency increases with the pool size as $\sim M^{b+\lambda \alpha}$. 
Also, the total power varies as $P(M) \sim \int S(f, M) df \sim M^{b + \lambda(\alpha -1)}$.

The numerically estimated critical exponents $b+\lambda\alpha = 2.67(2)$, $b = 0.77(1)$, and $b + \lambda(\alpha -1) = 1.68(2)$ [cf. Fig.~\ref{fig_dv_sf_2}] yield $\lambda = 0.99(4)$ and $\alpha = 1.9(1)$. As shown in the inset of Fig.~\ref{fig_dv_sf_1}, a clean data collapse well supports a precise estimation of the exponents within the statistical error. The previously reported numerical value for the spectral exponent $\alpha \approx 1.80(1)$~\cite{PhysRevLett.101.158702} does not differ significantly from our estimate of $\alpha = 1.9(1)$, although an exponent close to $\alpha \approx 2$ has a qualitatively different implication. As $N(t)$ shows a $1/f^2$-type spectrum, the avalanche size as a function of time $s(t)$ does not show temporal correlation. The estimated value of $\lambda \approx 1$ implies the cutoff frequency scales linearly with the system size. Although the primitive set size noise shows a trivial behavior of the $1/f^2$ form, the power spectra exhibit intriguing scaling features with the system size. Particularly, the power scales as $M^b$ in the non-trivial frequency regime.

\begin{figure}[t]
	\centering
	\scalebox{1}{\includegraphics{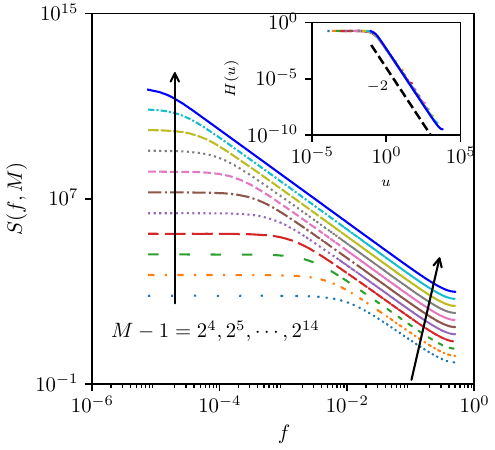}}
	\caption{L\'evy flight-A: (Main panel) The power spectra $S(f, M)$ for the random walk  $N(t)$ signals with different system sizes $M$. Here we consider lattice sites running from $-L$ to $L$ so that the total sites are $M = 2L+1$. The jump size is drawn form a power-law distribution $P(s) \sim 1/s^{2}$, truncated for $s\le L/2=M/4$. The linear extent is $L = 2^n$, where $n$ varies from 3 to 13. (Inset) The data collapse of the power spectra.}
	\label{fig_levy_1}
\end{figure}

\begin{table}[t]
\centering
\begin{tabular}{cccc}
\hline
\hline
~~~~~~Model ~~~~~~~~~& ~~~~~~~$\lambda$~~~~~~~ & ~~~~~~~$b$~~~~~~~ & ~~~~~~~$\alpha$~~~~~~~\\
\hline
Division model & 0.99(4) & 0.77(1) & 1.9(1)\\
L\'evy flight- A  & 0.97(2) & 1.02(1) & 2.00(6)\\
L\'evy flight- B &1.18(2) & 0.83(2) &1.97(8)\\
\hline
\hline
\end{tabular}
\caption{The numerically estimated critical exponents for power spectra properties.}
\label{tab1}
\end{table}

\begin{figure}[b]
	\centering
	\scalebox{1}{\includegraphics{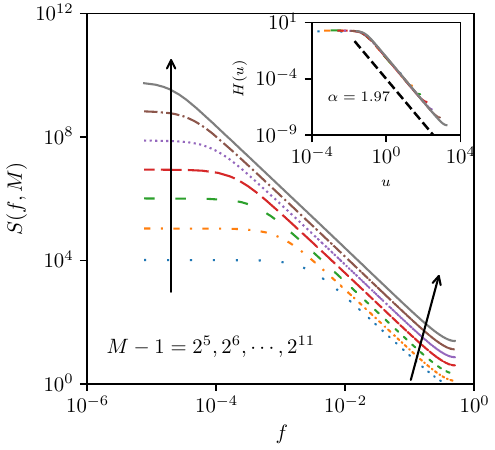}}
	\caption{Same as Fig.~\ref{fig_levy_1} for the L\'evy flight-B. The jump size is drawn form a power-law distribution $P(s) \sim 1/s^{2}$, truncated for  $s \le L/\ln_2{L} \sim M/\ln{M}$.}
	\label{fig_levy_2}
\end{figure}

\section{Power spectra of L\'evy flights}\label{sec_4}
To understand the stochastic nature of the primitive set size fluctuations, we examine a simple random walk with a power-law distributed jump size (L\'evy flight). Notice that L\'evy flights~\cite{vishwanathan_1999, VISWANATHAN2008133} have diverse application ranging from animal foraging patterns~\cite{Frederic_2005} to human movements~\cite{Brockmann_2006}. The random walk evolves in 1D with a range from $-L$ to $L$, such that the total lattice points are $M=2L+1$. We assume reflecting boundary conditions at $N = \pm (L+1)$ to ensure the random walk is bound. The random walk evolves as 
\begin{equation}
N(t+1)=N(t) + s,\nonumber
\end{equation} 
where $s$ is the jump size, randomly and independently drawn from a power-law probability distribution function $P(s) \sim s^{-(\tau'+1)}$ with $\tau'=1$. The walk starts from $N(t=0) = 0$. To set the mean position to zero or the random walk to remain symmetric, we take $s$ to be positive or negative with equal probability. 

We use the inverse sampling of the cumulative probability distribution function (CDF) method to generate the power-law distributed random numbers~\cite{PhysRevE.109.064135}. Let $\xi$ be a uniformly distributed random variable with a probability of 1 within the range $0\le \xi \le 1$ and 0 otherwise.
The general form of the CDF for the power-law distribution is often derived from the integration of PDF, resulting in $P_S(s) = 1-s^{-1}$ with $s_{min}=1$. Since the CDF is continuous, it will have a one-to-one mapping with numbers from the uniform distribution. Then, we find a nonlinear transformation $s = P_S^{-1}(\xi) = (1-\xi)^{-1}$.

In the division model, the avalanche size shows a cutoff size $\sim M/\ln{M}$. It is useful to set an upper limit for the jump size of the random walk. In this work, we have examined two cases. In the first case (say, L\'evy flight-A), the upper size limit is directly proportional to $M$ (linearly varying); more precisely, $s \le L/2 = M/4$. We generate the random walk signals via the Monte Carlo method and examine the power spectral density for such signals. As shown in Fig.~\ref{fig_levy_1}, the power spectra follow similar scaling features (the exponent $b$ differs slightly) to those discussed for the primitive set size noise in the division model. We apply the FSS method and estimate the critical exponents numerically: $b+\lambda\alpha = 2.96(1)$, $b = 1.02(1)$, $b+\lambda(\alpha-1)= 1.99(1)$, $\lambda = 0.97(2)$, and $\alpha = 2.00(6)$ [cf. Table~\ref{tab1}].

While the spectral and cutoff frequency exponents match well for the primitive set size in the division model and the random walk with power-law distributed jump limited linearly, the system size scaling exponent $b$ of the power in the non-trivial frequency regime is slightly different. We consider a second case where the jump size of the random walk is limited in a nonlinear manner as $s \le L/\ln_2{L} \sim M/\ln{M}$. We again compute the power spectra (cf. Fig.~\ref{fig_levy_2}) for the random walks for the second case (L\'evy flight-B) and perform the FSS analysis similarly. We find qualitatively similar scaling behavior. Table~\ref{tab1} presents a summary of the numerically estimated critical exponents. In this case, the spectral exponent $\alpha$ remains unchanged, and $b = 0.83(2)$ is closer to the division model. However, the cutoff frequency exponent $\lambda$ differs slightly.

\section{Summary}\label{sec_5}
We numerically studied the number theoretic division model of self-organized criticality. Given a set of $M-1$ ordered integers $\{2, 3, \cdots, M\}$, the model aims to form dynamically a primitive set of integers, where no integers can divide or be divisible by others. We examined the power spectra for the primitive set size fluctuations employing finite-size scaling and show that the spectrum trivially follows $1/f^2$-type scaling in the frequency regime $1/M \ll f \ll 1/2$ instead of, as reported previously, a spectrum with a scaling exponent significantly smaller than 2 (a signature of non-trivial temporal correlation, the $1/f$ noise). The $1/f^2$ spectrum in the primitive set size implies that the avalanches are temporally uncorrelated. Beyond this, our striking finding is that the power spectra, in the same frequency regime, also display a system size scaling, possibly of $\sim M^b$-type or proportional to the characteristic size $ \sim M/\ln{M}$ [both are indistinguishable because $b \approx 0.8$ and a plot of $S(f=0.3, M) \sim (M/\ln{M})^{\gamma}$ shows an exponent $\gamma =0.93(1)$]. Note that the cutoff frequency scales as $\sim 1/M$. We also found that a class of 1D random walk with a power-law distributed jump (L\'evy flight) can exhibit similar power spectral properties (quantitatively, the critical exponents differ slightly). We infer that the underlying dynamics of the primitive set size are superdiffusive.

\section*{ACKNOWLEDGMENTS}
RC would like to acknowledge UGC, India, for financial support through the Junior Research Fellowship. Through a seed grant under the IoE (Seed Grant-II/2022-23/48729), ACY acknowledges support from Banaras Hindu University. 

\bibliography{s1}
\bibliographystyle{apsrev4-1}

\end{document}